\documentclass[%
 reprint,
superscriptaddress,
preprintnumbers,
bibnotes,
amsmath,amssymb,
aps,
prl,
showkeys % Show keywords
]{revtex4-1}

\usepackage{amsmath}
\usepackage{graphicx}
\usepackage{url}
\usepackage{physics, braket}

\newcommand*{\citen}[1]{%
  \begingroup
    \romannumeral-`\x % remove space at the beginning of \setcitestyle
    \setcitestyle{numbers}%
    \cite{#1}%
  \endgroup   
}

\begin{document}

\title{General, Strong Impurity-Strength Dependence of Quasiparticle Interference}

\author{Seung-Ju Hong$^{\S}$}
\email{sjhong6230@snu.ac.kr}
\affiliation{Department of Physics and Astronomy, Seoul National University, Seoul 08826, Korea}

\author{Jae-Mo Lihm$^{\S}$}
\email{jaemo.lihm@gmail.com}
\affiliation{Department of Physics and Astronomy, Seoul National University, Seoul 08826, Korea}
\affiliation{Center for Correlated Electron Systems, Institute for Basic Science, Seoul 08826, Korea}
\affiliation{Center for Theoretical Physics, Seoul National University, Seoul 08826, Korea}

\author{Cheol-Hwan Park}
\email{cheolhwan@snu.ac.kr}
\affiliation{Department of Physics and Astronomy, Seoul National University, Seoul 08826, Korea}
\affiliation{Center for Correlated Electron Systems, Institute for Basic Science, Seoul 08826, Korea}
\affiliation{Center for Theoretical Physics, Seoul National University, Seoul 08826, Korea}

\begin{abstract}

Quasiparticle interference patterns in momentum space are often assumed to be independent of the strength of the impurity potential when comparing with other quantities, such as the joint density of states.
Here, using the $T$-matrix theory, we show that this assumption breaks down completely even in the simplest case of a single-site impurity on the square lattice with an $s$ orbital per site. Then, we predict from first principles a very rich, impurity-strength-dependent structure in the quasiparticle interference pattern of TaAs, an archetype Weyl semimetal.
This study thus demonstrates that the consideration of the details of the scattering impurity including the impurity strength is essential for interpreting Fourier-transform scanning tunneling spectroscopy experiments in general. 

\end{abstract}

\keywords{Quasiparticle interference, Scanning tunneling microscopy, Fourier-transform scanning tunneling spectroscopy, First-principles calculation, Wannier function, Impurity strength}

\maketitle

\section{Introduction}
Scanning tunneling microscopy plays a key role in nanoscience because it directly probes the surface topography and electronic density of states with a sub-nanometer resolution~\cite{binnig_scanning_1983,Shedd_1990}. 
In particular, the Fourier-transform scanning tunneling spectroscopy has been widely used to examine the surface electronic structures in momentum space~\cite{simon_fourier-transform_2011}. The experimental results are interpreted as the result of quasiparticle interference (QPI) induced by impurities or defects on the surface.
Theoretically, a QPI pattern is defined as the Fourier transform of the perturbation to the local density of states (LDOS) induced by the impurity~\cite{derry_quasiparticle_2015}.
The QPI patterns of various materials including Weyl semimetals~\cite{inoue_quasiparticle_2016,PhysRevB.93.035137,kourtis_universal_2016,PhysRevB.102.165117,rusmann_spin_2018}, high-$T_\mathrm{c}$ superconductors~\cite{PhysRevB.68.014508,kreisel_interpretation_2015,hoffman_imaging_2002,PhysRevB.67.020511,hirschfeld_robust_2015}, and topological insulators~\cite{rusmann_spin_2018,Rmann2020,yuan_electronic_2020} are being actively studied. 

One often analyzes the observed QPI patterns by comparing them to the joint density of states (JDOS)~\cite{simon_fourier-transform_2011}
\begin{equation}
    J(\mathbf{q};\omega) = \int\textup{d}\mathbf{k}\;\rho_0(\mathbf{k};\omega)\,\rho_0(\mathbf{k-q};\omega).
    \label{eq:JDOS}
\end{equation}
Here,
\begin{equation} \label{eq:DOS}
    \rho_0(\mathbf{k};\omega)=\frac{1}{\pi}\lim_{\eta\to 0^+}\textup{Tr Im } G_0(\mathbf{k};\omega-i\eta)
\end{equation}
is the surface density of states at wavevector {\bf k} and energy $\omega$ in the absence of impurities with $G_0$ the surface Green function.
Since there is no reference to the properties of the impurities in Eq.~\eqref{eq:JDOS}, the JDOS approximation neglects the impurity dependence of the QPI.

Another commonly used approximation of the QPI pattern is the spin scattering probability (SSP)~\cite{Roushan2009}
\begin{equation}
    J_s(\mathbf{q};\omega)=\sum_{i=0}^{3}\int\textup{d}\mathbf{k}\;\rho_i(\mathbf{k};\omega)\,\rho_i(\mathbf{k-q};\omega),
    \label{eq:SSP}
\end{equation}
where
\begin{equation}
    \rho_a(\mathbf{k};\omega)=\frac{1}{\pi}\lim_{\eta\to 0^+}\textup{Tr Im } \sigma_a\, G_0(\mathbf{k};\omega-i\eta)
\end{equation}
is the spin density along the $a$-th direction ($a=1,2,3$) with $\sigma_a$ the Pauli matrix.
Inoue \textit{et al.\,} used surface-projected SSP to explain the measured QPI patterns in TaAs~\cite{inoue_quasiparticle_2016}.
The SSP approximation goes beyond the JDOS approximation by forbidding the scatterings between oppositely-polarized pure spin states.
Still, SSP is independent of impurity-specific properties.

The Born approximation is also frequently used to simulate QPI. Under the Born approximation, one  takes into account only the first-order effect of the scattering potential on the Green function. Consequently, the QPI pattern computed within the Born approximation is independent of the strength of the scalar impurities up to an overall prefactor.
\begin{figure}[t]
    \centering
    \includegraphics[width=0.45\textwidth]{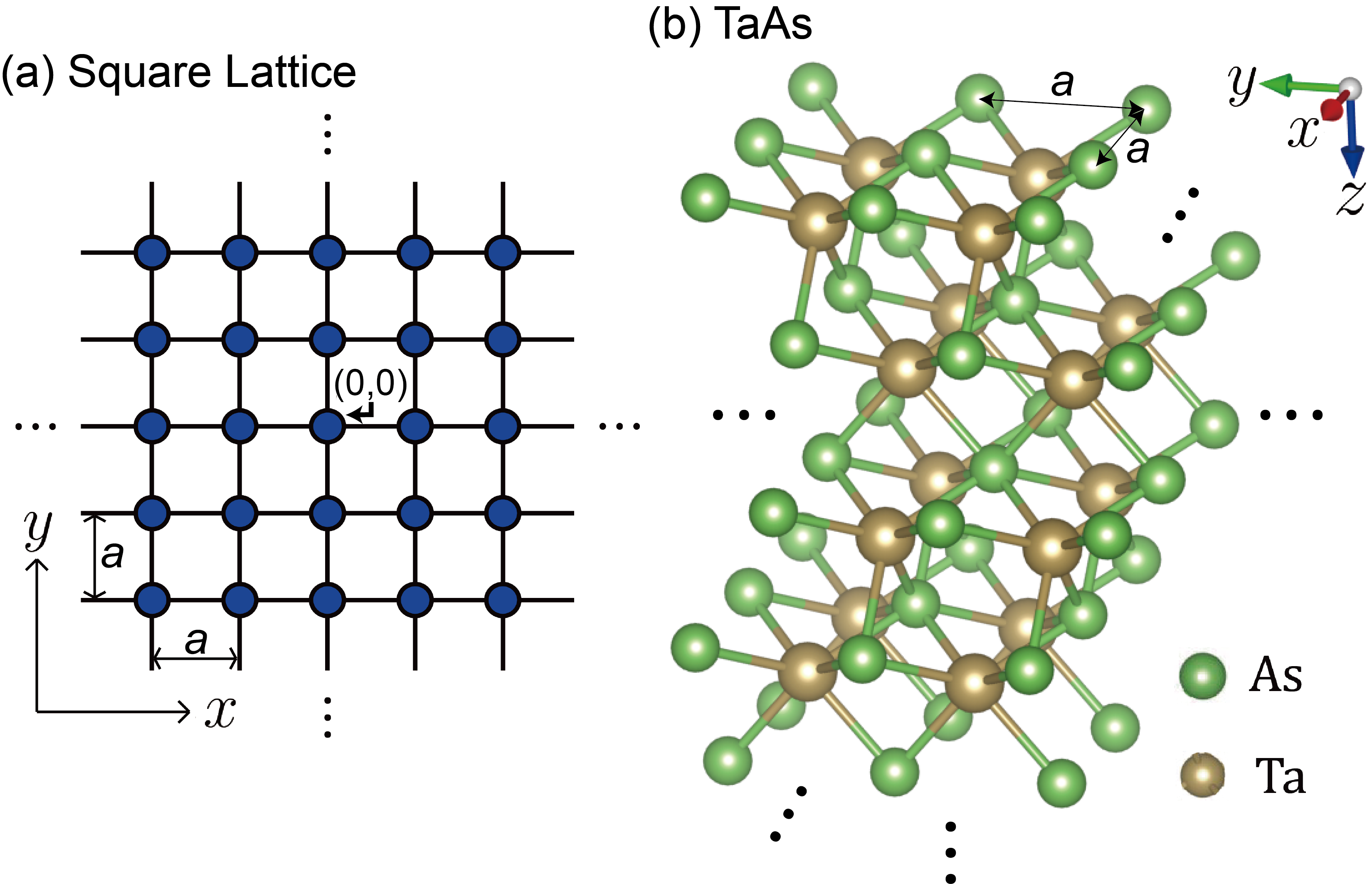}
    \caption{The structure of the systems examined in this paper. (a) Single-orbital square lattice. (b) As-terminated semi-infinite surface of TaAs.}
    \label{Systems}
\end{figure}

In this paper, we use the $T$-matrix method~\cite{Mahan2000} to investigate the effect of the strength of simple non-magnetic scalar impurities on the QPI pattern. Contrary to the common belief, the QPI pattern depends very sensitively on the impurity strength. Thus, the approximations that do not capture such impurity-strength dependence, such as JDOS, SSP, and the Born approximation, may fail to describe even the QPI arising from simple non-magnetic impurities.
We first show that the QPI patterns of a simple square lattice [Fig.~\ref{Systems}(a)] depends dramatically on the impurity strength. After fully understanding the physics of this simplest system, we look into TaAs [Fig.~\ref{Systems}(b)], an archetype Weyl semimetal, and find that its QPI pattern also depends significantly on the impurity strength.

\section{Methods}
The Green function of a semi-infinite surface in the absence of impurities is defined as
\begin{equation}
    \mathcal{G}_0(\omega) = (\omega-i0^+ -\mathcal{H}_0)^{-1},
\end{equation}
where $\mathcal{H}_0$ is the Hamiltonian of the pristine semi-infinite surface.
The Green function of the pristine surface is block diagonal with respect to the in-plane wavevectors.
The Green function for a wavevector $\bf{k}$ is
\begin{align}
\begin{split}
    G_0(\mathbf{k};\omega) = (\omega - i 0^+ - \mathcal{P}_\mathbf{k}\mathcal{H}_0\mathcal{P}_\mathbf{k})^{-1}
\end{split}
\end{align}
where $\mathcal{P}_\mathbf{k}$ is the projection operator onto the subspace with wavevector $\bf{k}$.

Now, let us introduce a scalar impurity with potential $\mathcal{V}$.
In this paper, we work using the tight-binding description of the Hamiltonian and consider the case where the impurity potential shifts the onsite energy of all orbitals of the topmost atom in the central in-plane unit cell.
Concretely, the impurity potential matrix element for orbitals $m$ and $n$ in in-plane unit cells $\mathbf{R}$ and $\mathbf{R'}$ is
\begin{equation} \label{eq:V_mel}
    \mathcal{V}_{m\mathbf{R},n\mathbf{R'}}
    = V (P_\textrm{T})_{m,n} \delta_{\mathbf{R},\mathbf{0}}\delta_{\mathbf{R'},\mathbf{0}},
\end{equation}
with $V$ the impurity strength.
Here, $P_\textrm{T}$ is a projection operator to the orbitals in the topmost atom, so that $(P_\textrm{T})_{m,n}$ is 1 if $m=n$ and $m$ is an orbital that belongs to the topmost atom, and 0 otherwise.

Using the $T$-matrix formalism, the change in the Green function induced by the perturbation $\mathcal{V}$ reads~\cite{simon_fourier-transform_2011}
\begin{equation} \label{eq:deltaG_Tmat}
    \Delta\mathcal{G}(\omega) = \mathcal{G}(\omega) - \mathcal{G}_0(\omega) = \mathcal{G}_0(\omega)\mathcal{T}(\omega)\mathcal{G}_0(\omega)
\end{equation}
where the $T$ matrix is defined as
\begin{equation}
    \mathcal{T}(\omega) = \mathcal{V}[\mathcal{I} - \mathcal{G}_0(\omega)\mathcal{V}]^{-1}\,,
    \label{T matrix}
\end{equation}
with $\mathcal{I}$ the identity operator.
For the impurity potential defined in Eq.~\eqref{eq:V_mel}, the $T$ matrix becomes
\begin{align} \label{eq:Tmat_onsite}
\begin{split}
    \mathcal{T}(\mathbf{R},\mathbf{R'};\omega)
    &= T\delta_{\bf{R},\bf{0}}\delta_{\mathbf{R'},\mathbf{0}} \\
    &=V\,[P_\textrm{T}-P_\textrm{T}\mathcal{G}_0(\mathbf{0,0};\omega)P_\textrm{T}V]^{-1}\delta_{\bf{R},\bf{0}}\delta_{\mathbf{R'},\mathbf{0}},
\end{split}
\end{align}
where the real-space Green function is defined as
\begin{equation}
    \mathcal{G}_0(\mathbf{R},\mathbf{R}';\omega) = \frac{1}{\Omega_{\textup{BZ}}}\int \textup{d}\mathbf{k}\; e^{-i\mathbf{k}\cdot(\mathbf{R}-\mathbf{R}')}G_0(\mathbf{k};\omega)
    \label{eq:Green_r}
\end{equation}
with $\Omega_{\textup{BZ}}$ the area of the 2D Brillouin zone.

To simulate scanning tunneling spectroscopy, we project the Green function to the topmost atomic layer, assuming that the only the LDOS of the topmost atomic layer is measured.
The change in the surface LDOS induced by the impurity reads
\begin{equation}
    \Delta \rho(\mathbf{R};\omega) = \frac{1}{\pi}\lim_{\eta\to 0^+}\textup{Im} \sum_{i\in\rm{topmost}} \left \langle w_{i\bf{R}}|\Delta \mathcal{G}(\omega-i\eta)|w_{i\bf{R}}\right \rangle,
\end{equation}
where $\ket{w_{i\bf{R}}}$ is a localized orthogonal basis function, such as the Wannier function.
The QPI pattern $\Delta \rho (\mathbf{q};\omega)$ is the Fourier transform of $\Delta \rho(\bf{R};\omega)$:
\begin{align} \label{QPI}
\begin{split}
    \Delta \rho (\mathbf{q};\omega)& =-\frac{i}{2\pi}\lim_{\eta\to 0^+}\sum_{i\in\rm{topmost}}
    \int\textup{d}\mathbf{k} \\ & \langle w_{i\bf{k-q}}|(\Delta\mathcal{G}(\omega-i\eta)-\Delta\mathcal{G}^{\dagger}(\omega-i\eta))|w_{i\bf{k}}\rangle.
\end{split}
\end{align}
If the system is invariant under a C$_2$ rotation with respect to $z$ [Fig.~\ref{Systems}], Eq.~(\ref{QPI}) reduces to
\begin{align} \label{eq:qpi_mirror}
\begin{split}
    \Delta \rho (\mathbf{q};\omega)
    &= \frac{1}{\pi}\lim_{\eta\to 0^+}\textup{Im}\sum_{i\in\rm{topmost}}
    \int\textup{d}\mathbf{k} \\ &\langle w_{i\bf{k-q}}|\Delta\mathcal{G}(\omega-i\eta)|w_{i\bf{k}}\rangle.
\end{split}
\end{align}
Using Eqs.~(\ref{eq:deltaG_Tmat}, \ref{eq:Tmat_onsite}, \ref{eq:qpi_mirror}) with the Green function of the pristine surface computed from the iterative method~\cite{sancho_highly_1985}, one can efficiently compute the QPI pattern induced by a localized potential impurity.

Now, as the simplest example, let us consider the square lattice with one $s$-like orbital per site with nearest-neighbor hopping with hopping integral $t$. The energy dispersion reads
\begin{equation}
    \epsilon_{\bf{k}} = 2t[\textup{cos}(ak_x)+\textup{cos}(ak_y)],
\end{equation}
where $a$ is the lattice parameter.
In this single-orbital case, the $T$ matrix becomes a complex number and so is $G_0(\mathbf{k}; \omega)$.
The QPI pattern becomes
\begin{equation}
    \Delta \rho (\mathbf{q};\omega) 
    = \frac{1}{\pi}\lim_{\eta\to 0^+} \textup{Im} \left[T\Pi(\mathbf{q};\omega-i\eta)\right]
    \label{eq:DeltaRhoTPi}
\end{equation}
where  
\begin{equation}
    \Pi(\mathbf{q};\omega) = \int\textup{d}\mathbf{k}\; G_0(\mathbf{k};\omega)\,G_0(\mathbf{k-q};\omega).
    \label{eq:Pi}
\end{equation}

\begin{figure*}[t]
    \centering
    \includegraphics[width=0.95\textwidth]{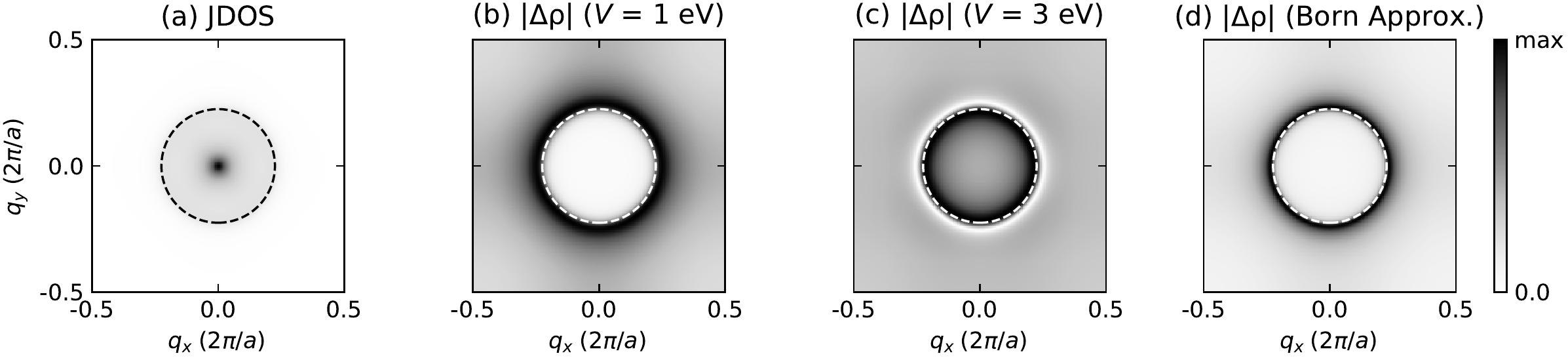}
    \caption{JDOS [(a)], absolute value of the QPI pattern [(b) and (c)], and absolute value of the QPI pattern computed within the Born approximation [(d)] for the single-orbital square lattice with $t=-1$~eV at $\omega=3.5$~eV. The dashed curves are circles with radius $q=2k_{\rm c}$.}
    \label{JDOS_QPI_of_square_lattice}
\end{figure*}

\section{Results and Discussion}
Figure~\ref{JDOS_QPI_of_square_lattice}(a) and (b, c) show the computed JDOS and QPI patterns of the single-orbital square lattice with $t$=-1~eV, respectively. The quantities are calculated at $\omega = 3.5$~eV. At this energy, the constant-energy contour is approximately a circle with radius $k_\textup{c} = 0.225\pi / a$. Comparing the JDOS and the QPI patterns, we find that the intensity outside the $q=2k_{\rm c}$ circle are clearly different: the JDOS is zero outside the $q=2k_{\rm c}$ circle, while the QPI patterns are non-zero.
More importantly, we find a large difference between the QPI patterns for $V = 1$~eV and $V = 3$~eV.

Within the Born approximation, the $T$ matrix is replaced by the impurity potential $V$.
Therefore, the computed QPI pattern becomes independent of the impurity strength $V$ except for a proportionality constant.
The QPI pattern computed within the Born approximation [Fig.~\ref{JDOS_QPI_of_square_lattice}(d)] is more similar to the exact QPI pattern for small $V$ [Fig.~\ref{JDOS_QPI_of_square_lattice}(b)] than that for large $V$ [Fig.~\ref{JDOS_QPI_of_square_lattice}(c)] since the Born approximation becomes more accurate for weaker perturbations.
However, the QPI pattern of a stronger impurity [Fig.~\ref{JDOS_QPI_of_square_lattice}(c)] considerably deviate from the Born approximation.
\begin{figure}[t]
    \centering
    \includegraphics[width=0.45\textwidth]{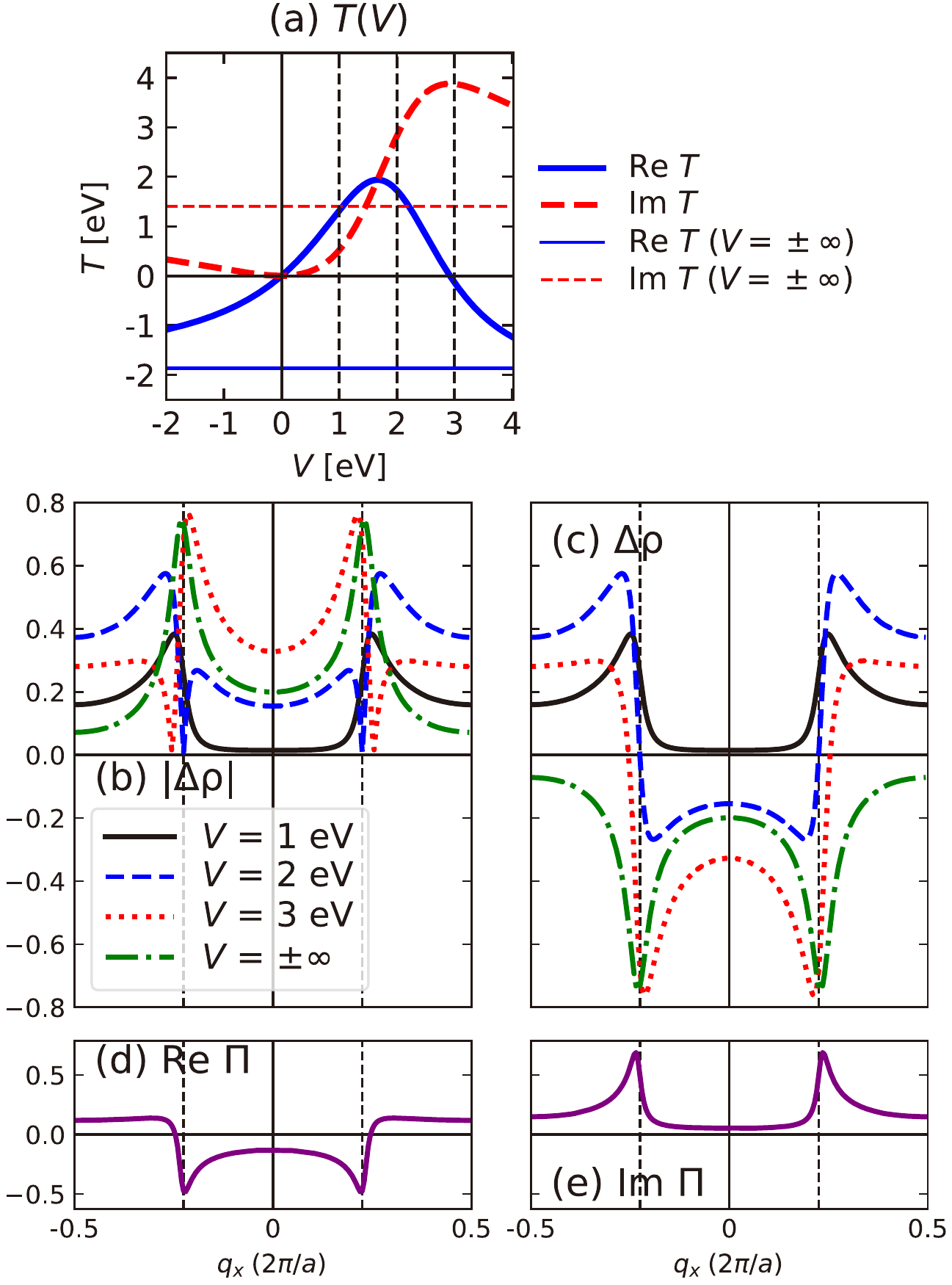}
    \caption{Real and imaginary parts of $T$ [Eq.~\eqref{eq:Tmat_onsite}] versus $V$ [(a)]. Absolute [(b)] and signed [(c)] values of the QPI patterns for various values of $V$. Real [(d)] and imaginary [(e)] parts of $\Pi$ [Eq. (\ref{eq:Pi})]. All the results are for the single-orbital square lattice with $t=-1$~eV at $\omega=3.5$~eV. The dashed vertical lines in (b)-(e) indicate $q=2k_\mathrm{c}$. We note that $\Delta \rho(\mathbf{q};\omega)$ is real valued [Eq. (\ref{eq:qpi_mirror})].}
    \label{fig:T_QPI_PI}
\end{figure}

To understand the impurity-strength dependence of the QPI patterns shown in Fig.~\ref{JDOS_QPI_of_square_lattice}, we examine the $T$ matrix. Figure~\ref{fig:T_QPI_PI}(a) shows the real and imaginary parts of $T$ as a function of $V$. Also, we plot $T$ at $V = \pm \infty$, which can be used to simulate a vacancy at the impurity site. At small $|V|$, the real part of $T$ can be approximated as $V$ and its imaginary part is negligible, indicating the adequacy of the Born approximation. However, at large $\abs{V}$, both $\Re T$ and $\Im T$ are sizable.
The QPI signal is proportional to the imaginary part of $T\Pi$ [Eq.~\eqref{eq:DeltaRhoTPi}].
Thus, only the imaginary part of $\Pi$ contribute to the QPI pattern at small $|V|$, while both $\Im \Pi$ and $\Re \Pi$ contributes at larger $\abs{V}$.

Using the obtained $T$ matrix, we now aim to understand the $V$ dependence of the QPI patterns [Fig.~\ref{JDOS_QPI_of_square_lattice}(b) and (c)].
In Fig.~\ref{fig:T_QPI_PI}(b), we show the absolute value of the QPI patterns along the line $q_y =0$.
This seemingly complicated dependence of the QPI pattern on $V$ can be simply understood by comparing the signed QPI patterns [Fig.~\ref{fig:T_QPI_PI}(c)] and the $\Pi$ function [Eq.~\eqref{eq:Pi} and Figs.~\ref{fig:T_QPI_PI}(d,e)] (see also Fig.~S1~\cite{SuppInfo}).
The complementary feature of $\Re \Pi$ and $\Im \Pi$ is the key to understanding the impurity-strength dependence of the QPI pattern. Using Eq.~\eqref{eq:DeltaRhoTPi}, one can write the QPI signal as
\begin{align}
\begin{split}
    \Delta\rho(\mathbf{q};\omega)
    &= \frac{1}{\pi}\lim_{\eta\to 0^+}[
    \Im T \; \Re \Pi(\mathbf{q};\omega-i\eta)
    \\&+ \Re T \; \Im \Pi(\mathbf{q};\omega-i\eta)].
\end{split}
\label{Re and Im Pi}
\end{align}
Apparently, the QPI signal is a linear combination of $\Re \Pi$ and $\Im \Pi$ with coefficients $-\frac{1}{\pi}\Im T$ and $-\frac{1}{\pi}\Re T$, respectively.
For $V = 1$~eV, since $|{\rm Im}~T|<|{\rm Re}~T|$ [Fig.~\ref{fig:T_QPI_PI}(a)], $\Im \Pi$ dominates the qualitative features of the QPI pattern.
However, for $V = 3$~eV, we find $|{\rm Im}~T|\gg|{\rm Re}~T|$ [Fig.~\ref{fig:T_QPI_PI}(a)], and thus the QPI pattern is almost exclusively determined by $\Re \Pi$.
This analysis clearly demonstrates why the signed QPI patterns for $V=1$~eV and $V=3$~eV [Fig.~\ref{fig:T_QPI_PI}(c)] resemble $-{\rm Im}~\Pi$ [Fig.~\ref{fig:T_QPI_PI}(e)] and $-{\rm Re}~\Pi$ [Fig.~\ref{fig:T_QPI_PI}(d)], respectively. In brief, the position of peaks are determined by $\Pi$ or the Green function $G_0(\mathbf{k}; \omega)$ [Eq.(\ref{eq:Pi})], while their intensity and shape are determined by $T$.

In passing, we note that Ref.~\citen{PhysRevB.84.134507} investigated the impurity-strength dependence of the QPI patterns of the parent compounds of iron-pnictide superconductors, magnetic materials with a spin density wave order. The subjects and findings of our work and Ref.~\citen{PhysRevB.84.134507} are totally different as in that work, the peaks in the energy position of the spin-density wave are of crucial importance. We also note that the findings in our work are relevant to general, even non-magnetic systems.

In order to see the effect of real-space masking performed as a post-processing of the experimental data~\cite{Rmann2020}, we applied the real-space masking to our calculated QPI patterns. We find that the real-space masking still preserves much of the impurity-strength dependence of the QPI patterns (see Fig.~S2~\cite{SuppInfo}).

We now move on to the As-terminated surface of TaAs. We used the \texttt{Quantum ESPRESSO} package~\cite{giannozzi_quantum_2009,giannozzi_advanced_2017} for density-functional theory calculations and the Wannier90 package~\cite{mostofi_updated_2014,Pizzi2020} to construct the \textit{ab initio} Wannier-function-based tight-binding models. (See Supplementary Information for the computational details~\cite{SuppInfo}.)
When computing the JDOS [Eq.~\eqref{eq:JDOS}], the SSP [Eq.~\eqref{eq:SSP}], and the QPI pattern [Eq.~\eqref{QPI}], the sum over atomic orbitals was limited to the orbitals belonging to the topmost As atoms.

\begin{figure*}[t]
    \centering
    \includegraphics[width=\textwidth]{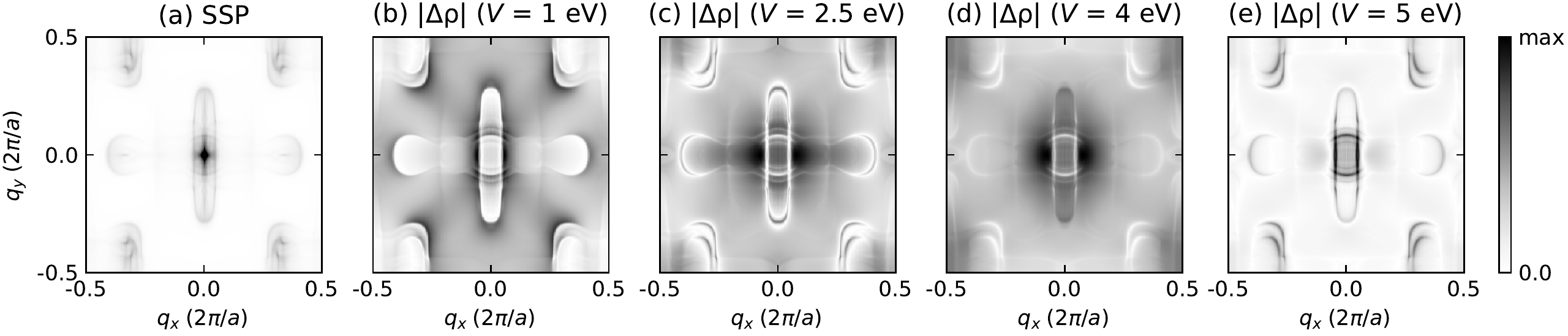}
    \caption{SSP [(a)], and the absolute value of the QPI patterns [(b)-(e)] for the As-terminated TaAs surface with a single on-site impurity for $\omega$ at the Fermi level. $V$ represents the on-site potential shift of the surface As p orbitals.}
    \label{JDOS SSP QPI of TaAs}
\end{figure*}

The SSP and the QPI patterns of TaAs are shown in Fig.~\ref{JDOS SSP QPI of TaAs}. The JDOS (see Fig. S3 and S4~\cite{SuppInfo}) and SSP have only minor differences~\cite{inoue_quasiparticle_2016}. However, as in the case of the square lattice (Fig.~\ref{JDOS_QPI_of_square_lattice}), the QPI pattern is widely different from the JDOS and SSP.
Moreover, the calculated QPI patterns are strongly dependent on the impurity strength.

To understand this strong impurity-strength dependence of the QPI patterns, we investigate the corresponding signed QPI signals. Figure S5~\cite{SuppInfo} shows that the positions of the peaks of $\Delta \rho$ do not strongly depend on $V$ and are determined by the electronic structures of TaAs, as similar peaks also occur in the SSP (or JDOS; see Fig.~S4~\cite{SuppInfo}). However, the signed QPI patterns vary strongly with $V$. These behaviors are very similar to the case of the square lattice [Fig.~\ref{fig:T_QPI_PI}(c)].
Since the impurity potential acts on multiple orbitals of the topmost As atom, the $T$ matrix is now a matrix, not a complex number. Hence, a simple analysis like Eq.~\eqref{Re and Im Pi} is not possible. However, as in the square lattice case, the position of the peaks are mainly determined by the pristine Green functions. Also, the impurity-strength dependence of the $T$ matrix is the origin of the complex impurity-strength dependence of the QPI patterns shown 
clearly in Fig.~\ref{JDOS SSP QPI of TaAs}.

In reality, there are usually many kinds of impurities in the Fourier-transformed region. Thus, it is necessary to take account of multi-impurity effects in QPI patterns. To deal with multiple impurities, one must consider two distinct factors: multiple scatterings from different impurities, and different positions of the impurities. For the first part, we note first that multiple scatterings from a single isolated impurity are well taken care of by the $T$-matrix method. The assumption of neglecting multiple scatterings from different impurities is well accepted by the community if the density of impurities is not too high~\cite{PhysRevB.91.235127,PhysRevB.67.020511,Rmann2020}.

For the second point, the effect of the different positions of multiple impurities can be taken into account by considering the phase factor $e^{i\mathbf{q}\cdot\mathbf{R}_i}$, where $\mathbf{R}_i$ is the position of the \textit{i}-th impurity. For example, in Ref.~\citen{Sharma2021}, the authors considered the impurity position dependence of the QPI patterns by multiplying this phase factor $e^{i\mathbf{q}\cdot\mathbf{R}_i}$ to the single impurity QPI pattern.

When we calculate the theoretical QPI patterns, we must reflect the positions of experimentally observed impurities. For multiple impurities, the resultant QPI pattern is given as $\sum_i e^{i\mathbf{q}\cdot\mathbf{R}_i} \Delta\rho_i$ where $\Delta\rho_i$ denotes the QPI pattern if only the \textit{i}-th impurity was present at the origin. This result is a simple sum of the terms arising from a single impurity since the inter-impurity scatterings can be safely neglected in most experimentally relevant cases~\cite{PhysRevB.91.235127,PhysRevB.67.020511,Rmann2020}. We then can compare this QPI pattern or its real-space Fourier transform with the experimental STM/FT-STS result.

Furthermore, in real experiments, we can break down the overall QPI patterns into contributions arising from each isolated impurity. For example, one can approximate $\Delta \rho_i(\mathbf{r})$ with a parametric form with a small number of parameters and fit $\sum_i \Delta \rho_i(\mathbf{r} - \mathbf{R}_i)$ to the experimental STM data. Note that $\Delta \rho_i(\mathbf{r})$ will be the same for the same kind of impurities.

\begin{figure*}[!tb]
    \centering
    \includegraphics[width=\textwidth]{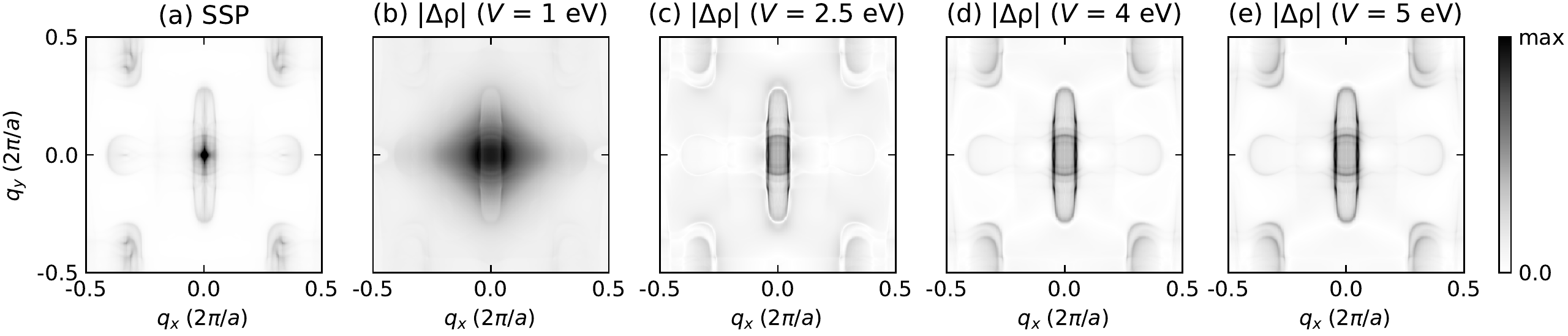}
    \caption{SSP [(a)], and the absolute value of the QPI patterns [(b)-(e)] for the As-terminated TaAs surface for $\omega$ at the Fermi level. Here, the on-site impurity potential at the p orbitals of strength $V$ was applied at a single surface As atom. The hopping integral from the p orbitals of that atom to the p orbitals of the adjacent As atoms was altered by 0.5$V$.}
    \label{fig:nn_hopping}
\end{figure*}

Although we restricted the theoretical discussion to $\mathbf{R}=\mathbf{R}'=\mathbf{0}$ case for simplicity, the formalism can be extended to the case with non-zero impurity matrix elements for $\mathbf{R}\neq\mathbf{0}$ or $\mathbf{R}'\neq\mathbf{0}$.  We calculate $T(\mathbf{R},\mathbf{R'};\omega)$ with Eq.~\eqref{eq:Green_r}. If the range of $V$ is finite in real space, the matrix equation for the $T$-matrix is a finite-dimensional matrix algebra and can be easily solved numerically. Then, we perform the sum over $\mathbf{R}$ and $\mathbf{R}'$ for the computation of QPI patterns. We performed this computation for an impurity that alters the nearest-neighbor hopping and for an impurity that shifts the on-site potential of nearest-neighboring atoms. The results are shown in Figs.~\ref{fig:nn_hopping} and S6~\cite{SuppInfo}, respectively. Again, we can see that the QPI patterns depend much on the properties of the impurity.

We now provide an intuitive explanation for the discrepancy between JDOS, SSP, Born approximation and the $T$-matrix formalism. We first concentrate on the difference between JDOS/SSP and the Born approximation. For the single-orbital case, the trace in Eq.~(\ref{eq:DOS}) reduces to single term and the density of states is equal to the imaginary part of the Green function. Thus, Eq.~(\ref{eq:JDOS}) can be written as
\begin{equation}
    J(\mathbf{q};\omega) = \frac{1}{\pi^2} \int\textup{d}\mathbf{k}\;\textup{Im}\,G_0(\mathbf{k};\omega)\,\textup{Im}\,G_0(\mathbf{k-q};\omega).
    \label{eq:JDOS_Green}
\end{equation}
(We omitted $\lim_{\eta\to 0^+}$ for brevity; the limit is always implied.) The equation for QPI under the Born approximation is obtained from Eq.~\eqref{eq:DeltaRhoTPi}, and the result is
\begin{widetext}
\begin{align}
\begin{split}
    \Delta\rho(\mathbf{q};\omega) &= \frac{V}{\pi} \int\textup{d}\mathbf{k}\;\textup{Im}\,[G_0(\mathbf{k};\omega)\,G_0(\mathbf{k-q};\omega)] \\
    & = \frac{V}{\pi}
    \left [ \int\textup{d}\mathbf{k}\;\textup{Re}\,G_0(\mathbf{k};\omega)\,\textup{Im}\,G_0(\mathbf{k-q};\omega) + \int\textup{d}\mathbf{k}\;\textup{Im}\,G_0(\mathbf{k};\omega)\,\textup{Re}\,G_0(\mathbf{k-q};\omega)\right ].
    \label{eq:QPI_Green}
\end{split}
\end{align} 
\end{widetext}
If the system is invariant under a C$_2$ rotation with respect to $z$, we have
\begin{equation}
    \Delta\rho(\mathbf{q};\omega) = \frac{2T(\omega)}{\pi} \int\textup{d}\mathbf{k}\;\textup{Re}\,G_0(\mathbf{k};\omega)\,\textup{Im}\,G_0(\mathbf{k-q};\omega). 
    \label{eq:QPI_Green_C2}
\end{equation}
From Eq.~(\ref{eq:JDOS_Green}) and Eq.~(\ref{eq:QPI_Green_C2}), we see that the real part of the Green function contributes to the QPI intensity within the Born approximation. Due to this real part of the Green function, scatterings from and to states which are not on the Fermi surface also contribute to QPI as opposed to JDOS.~\cite{PhysRevB.93.035137}

We can further simplify Eq.~\eqref{eq:QPI_Green_C2} by using the Kramers-Kronig relation between $\textup{Re}\,G_0(\mathbf{k};\omega)$ and $\textup{Im}\, G_0(\mathbf{k};\omega)$:
\begin{align}
\begin{split}
    \Delta\rho(\mathbf{q};\omega) &= \frac{2T(\omega)}{\pi} \int\textup{d}\mathbf{k}\;\textup{d}\omega'\;\frac{1}{\omega'-\omega} \\ &\times\textup{Im}\,G_0(\mathbf{k};\omega')\,\textup{Im}\,G_0(\mathbf{k-q};\omega). 
    \label{eq:QPI_Green_kk_relation}
\end{split}
\end{align}
Thus, the transitions involving states with energy $\omega'$ different from $\omega$ participate in the QPI but do not contribute to the JDOS [Eq.~(\ref{eq:JDOS_Green})] and SSP [Eq.~\eqref{eq:SSP}].

Equation \eqref{eq:QPI_Green_kk_relation} holds for general impurity potentials. Using the Born approximation to take the $V\rightarrow 0$ limit, we find
\begin{align}
\begin{split}
    \Delta\rho(\mathbf{q};\omega) &= \frac{2V}{\pi} \int\textup{d}\mathbf{k}\;\textup{d}\omega'\;\frac{1}{\omega'-\omega}\\ & \times\textup{Im}\,G_0(\mathbf{k};\omega')\,\textup{Im}\,G_0(\mathbf{k-q};\omega).
    \label{eq:QPI_JDOS}
\end{split}
\end{align}
Comparing Eq.~\eqref{eq:QPI_JDOS} with JDOS [Eq.~\eqref{eq:JDOS_Green}], we can see that the off-shell states are included in the QPI patterns [$\Delta\rho(\mathbf{q};\omega)$] even in the limit of $V\rightarrow 0$, while only the on-shell states contribute to JDOS.  Thus, JDOS (or SSP) is not the correct $V\rightarrow 0$ limit of QPI patterns~\cite{derry_quasiparticle_2015,PhysRevB.91.235127}, and we must always consider the off-shell states to correctly interpret QPI patterns. This point is visually demonstrated in Fig.~\ref{JDOS_QPI_of_square_lattice}(a) and (d). The JDOS [Fig.~\ref{JDOS_QPI_of_square_lattice}(a)] is non-zero only inside the circle in momentum space with radius $q=2k_{c}$, while the Born approximation QPI intensity [Fig.~\ref{JDOS_QPI_of_square_lattice}(d)] is higher outside the circle.

Next, let us compare the $T$-matrix formalism and the Born approximation.
The $T$-matrix $T(\omega) = V[I-G_0(\omega)V]^{-1}$ is a sensitive function of $V$, especially near the pole of the Green function. This sensitivity gives the strong dependence of QPI patterns on the impurity strength. The frequency-dependence of $T(\omega)$ arises from multiple scattering effects as we can easily see from the series expansion $T(\omega) = V + VG_0(\omega)V + \cdots$.
Such an effect is ignored in the Born approximation because only the leading order term is taken: $T(\omega) \approx V$.

One might argue that what one cares about in the QPI patterns is just the scattering wave vectors, and thus, JDOS, SSP, or Born approximation can give a reliable description of the QPI patterns. However, extracting scattering wave vectors from QPI patterns may not be trivial or straightforward in some cases so that simple approximations can fail to describe even the scattering wave vectors. For example, let us pay attention to the regions near $\mathbf{q}=0$ in Fig.~\ref{fig:nn_hopping}. In the case of Fig.~\ref{fig:nn_hopping}(b), some scattering vectors are obscured by the broad intensity distribution centered at $\mathbf{q}=0$. Also, the scattering vectors at the corners are weakened. It is obvious to see that extracting the scattering vectors is not always unambiguous and it is quite important to be able to understand the QPI patterns themselves.

\begin{figure}
    \centering
    \includegraphics[width=0.45\textwidth]{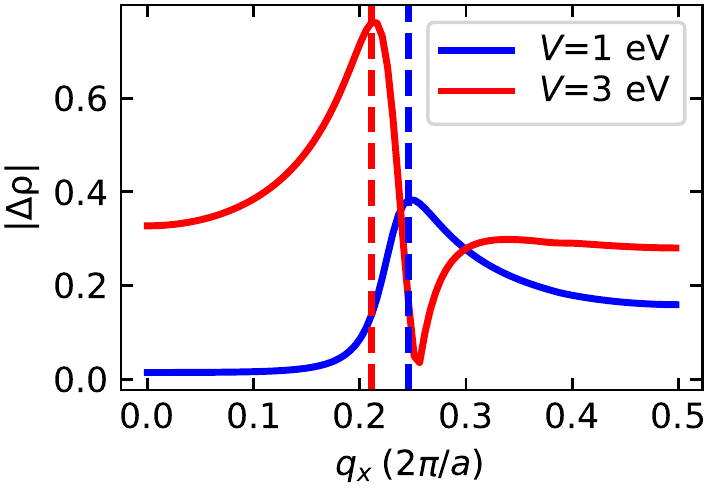}
    \caption{Absolute value of the QPI pattern along $q_y=0$ which is shown in Fig.~\ref{JDOS_QPI_of_square_lattice}(b) and (c) for the 2D square lattice with an s-like orbital per site}
    \label{fig:max_intensity}
\end{figure}

Furthermore, if we look at Fig.~\ref{fig:max_intensity}, we can find that even in the simplest case of the square lattice with an $s$ orbital per site, the maximum intensity scattering wave vectors for $V =1$~eV and $V=3$~eV are different. The difference is $\Delta q=0.070\pi/a$, while the scattering vector from the density of states is $q_c=2k_c=0.450\pi/a$. The relative difference in the scattering wave vector is 16 \%, which is non-negligible. Therefore, for an accurate extraction of the scattering wave vectors, we must consider the impurity strength dependence.

Moreover, people usually Fourier transform the STM pattern over a large area which necessarily contains different kinds of impurities. For example, Fig. 1B of Ref.~\citen{inoue_quasiparticle_2016} shows several different kinds of impurities in the topographic image. The $\textup{d}I/\textup{d}V$ image for this surface is shown in Fig. 2D of Ref.~\citen{inoue_quasiparticle_2016}. If we Fourier transform this pattern, the QPI patterns with different impurities will mix up. However, as described in our manuscript, the QPI pattern is strongly dependent on the impurities. For a more accurate investigation of the QPI patterns, one must decompose individual QPI patterns into contributions arising from each impurity.

Although the topographic images in Fig. 1B of Ref.~\citen{inoue_quasiparticle_2016} shows that there are three different types of impurities with different scattering properties, this information is not sufficient to pinpoint the impurities from first-principles calculations.  We need more information on the chemistry of the surface condition and the impurities which can be obtained from experiences or further experiments, not necessarily confined to scanning tunneling spectroscopy.

As a side note, let us compare the reported experimental and theoretical QPI patterns, Fig. 4H and Fig. 4K of Ref.~\citen{inoue_quasiparticle_2016}. There, the intensity maps around $\mathbf{q}=0$ are quite different from each other: it is clear that a simple Gaussian or Lorentzian broadening of the theoretical spectrum cannot reproduce the experimental one. Furthermore, if we concentrate on the corners, e.g., near ($q_x, q_y$)=($2\pi/a, 2\pi/a$), we see that the QPI pattern is absent in the theoretical spectrum unlike in the experimental one. This is due to the neglect of Umklapp scatterings in the theoretical spectrum of {Ref.~\citen{inoue_quasiparticle_2016}}. In TaAs, a good part of surface states are localized near the Brillouin zone boundary (see Fig. S3~\cite{SuppInfo}). Hence, the small-$|\mathbf{q}|$ Umklapp ($\mathbf{q+G}$) scattering is important.
These factors are not directly related to the impurity-strength dependence of QPI patterns.
Still, they reflect the fact that the current theoretical understanding of QPI patterns is rather incomplete. 

The above discussions show that a complete analysis of QPI patterns is far from trivial. To reach an agreement between theory and experiment, we need (i) the information on at least roughly what kind(s) of impurities are present, and (ii) first-principles calculations starting from such knowledge. Our work provides the tool for the latter part (a first-principles method) for the understanding of QPI patterns. However, we still need the former part, the experimental information on the impurities. Required first-principles calculations will depend on the quantity and quality of the experimentally obtained information on impurities. By showing the impurity-strength dependence of the QPI spectra, our work informs QPI experimentalists of the importance of impurities and may lead them to conduct experiments on the types of impurities and the dependence of QPI spectra on the impurities, in systems not limited to TaAs. Then, using those further experimental data and the $T$-matrix method based on first-principles calculations, people may fully understand the physics of FT-STS and QPI.

\section{Conclusion}
In conclusion, we have shown that the QPI pattern strongly depends on the scattering properties of impurities, including its strength.
This finding holds even in the simplest case of the square lattice with non-magnetic, scalar onsite impurities.
The impurity-strength dependence is also present in the QPI patterns of TaAs. We were able to describe the complex impurity-strength dependence of the QPI patterns of both systems from a unified framework: the pristine surface Green functions determines the position of the peaks, while the intensity and shape of the peaks are significantly affected by the $T$ matrix.
Our findings that the QPI patterns can be completely different for different types of impurities are general, with their profound applicability ranging from the simplest toy model to complicated topological materials, and thus, it is impossible to fully understand the QPI patterns without reference to the types of impurities in QPI experiments. Therefore, our findings should generally be used in analyzing the results of Fourier-transform scanning tunneling spectroscopy experiments.

\section{Supplementary Information}
\noindent{$\bullet$ Details of first-principles calculations and Green function calculation}

\noindent{$\bullet$ Figures and explanation about real-space masking}

\noindent{$\bullet$ Fourier-transformed LDOS and surface projected JDOS for As-terminated TaAs surface}

\noindent{$\bullet$ Signed values of the QPI signal for the As-terminated TaAs surface along different paths in the Brillouin zone}

\noindent{$\bullet$ SSP and the absolute value of QPI patterns for the As-terminated TaAs surface with different impurity potential values}

\begin{acknowledgments}
\section{Acknowledgement}
$^\S$S.-J.H. and J.-M.L. contributed equally to this work. This work was supported by the Creative-Pioneering Research Program through Seoul National University, Korean NRF No-2020R1A2C1014760, and the Institute for Basic Science (No. IBSR009-D1). Computational resources were provided by KISTI Supercomputing Center (KSC-2019-CRE-0246).
\end{acknowledgments}

\bibliography{ms}

\end{document}

\title{General, Strong Impurity-Strength Dependence of Quasiparticle Interference: \\Supplementary Information}

\author{Seung-Ju Hong$^{\S}$}
\email{sjhong6230@snu.ac.kr}
\affiliation{Department of Physics and Astronomy, Seoul National University, Seoul 08826, Korea}

\author{Jae-Mo Lihm$^{\S}$}
\email{jaemo.lihm@gmail.com}
\affiliation{Department of Physics and Astronomy, Seoul National University, Seoul 08826, Korea}
\affiliation{Center for Correlated Electron Systems, Institute for Basic Science, Seoul 08826, Korea}
\affiliation{Center for Theoretical Physics, Seoul National University, Seoul 08826, Korea}

\author{Cheol-Hwan Park}
\email{cheolhwan@snu.ac.kr}
\affiliation{Department of Physics and Astronomy, Seoul National University, Seoul 08826, Korea}
\affiliation{Center for Correlated Electron Systems, Institute for Basic Science, Seoul 08826, Korea}
\affiliation{Center for Theoretical Physics, Seoul National University, Seoul 08826, Korea}

\maketitle

\beginsupplement
\section{Green Function Calculation}
We applied the iterative scheme to obtain the surface Green function for the pristine semi-infinite surface~\cite{sancho_highly_1985}.
For the calculation of the QPI patterns, in the case of the single-orbital square lattice, we used a $200\times200$ k-point grid and a 0.05~eV broadening.
In the case of TaAs surface, we used a $400\times400$ k-point grid and a 0.01~eV broadening.

\section{Real-space Masking}

\begin{figure}[!tb]
    \centering
    \includegraphics[width=0.45\textwidth]{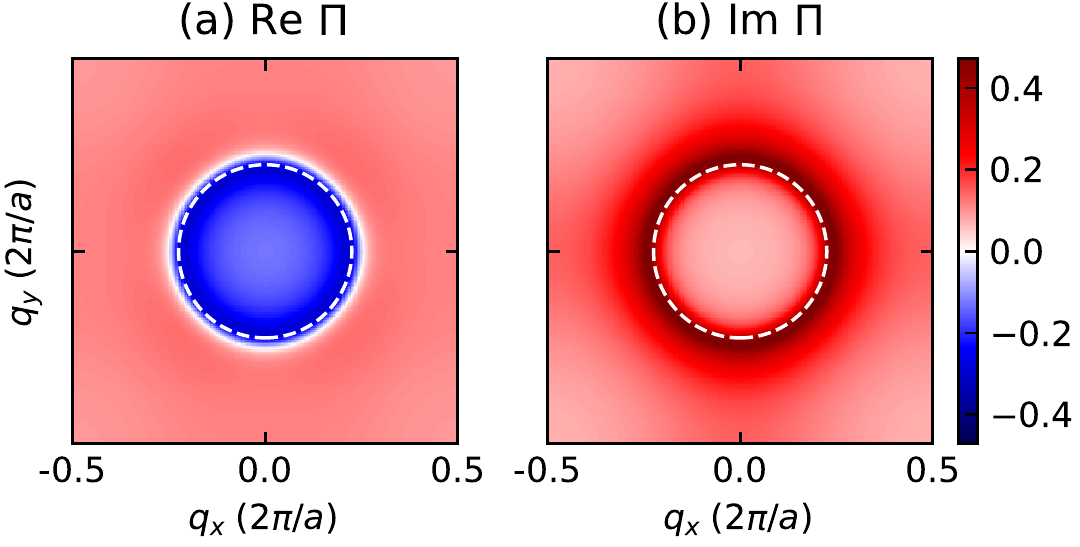}
    \caption{Real [(a)] and imaginary [(b)] parts of $\Pi$ [Eq.~(18) of the main manuscript] for the square lattice with $t=-1$~eV at $\omega=3.5$~eV. The dashed curves are circles with radius $q=2k_{\rm c}$.}
    \label{fig:supp_Pi}
\end{figure}

\begin{figure*}[!tb]
    \centering
    \includegraphics[width=\textwidth]{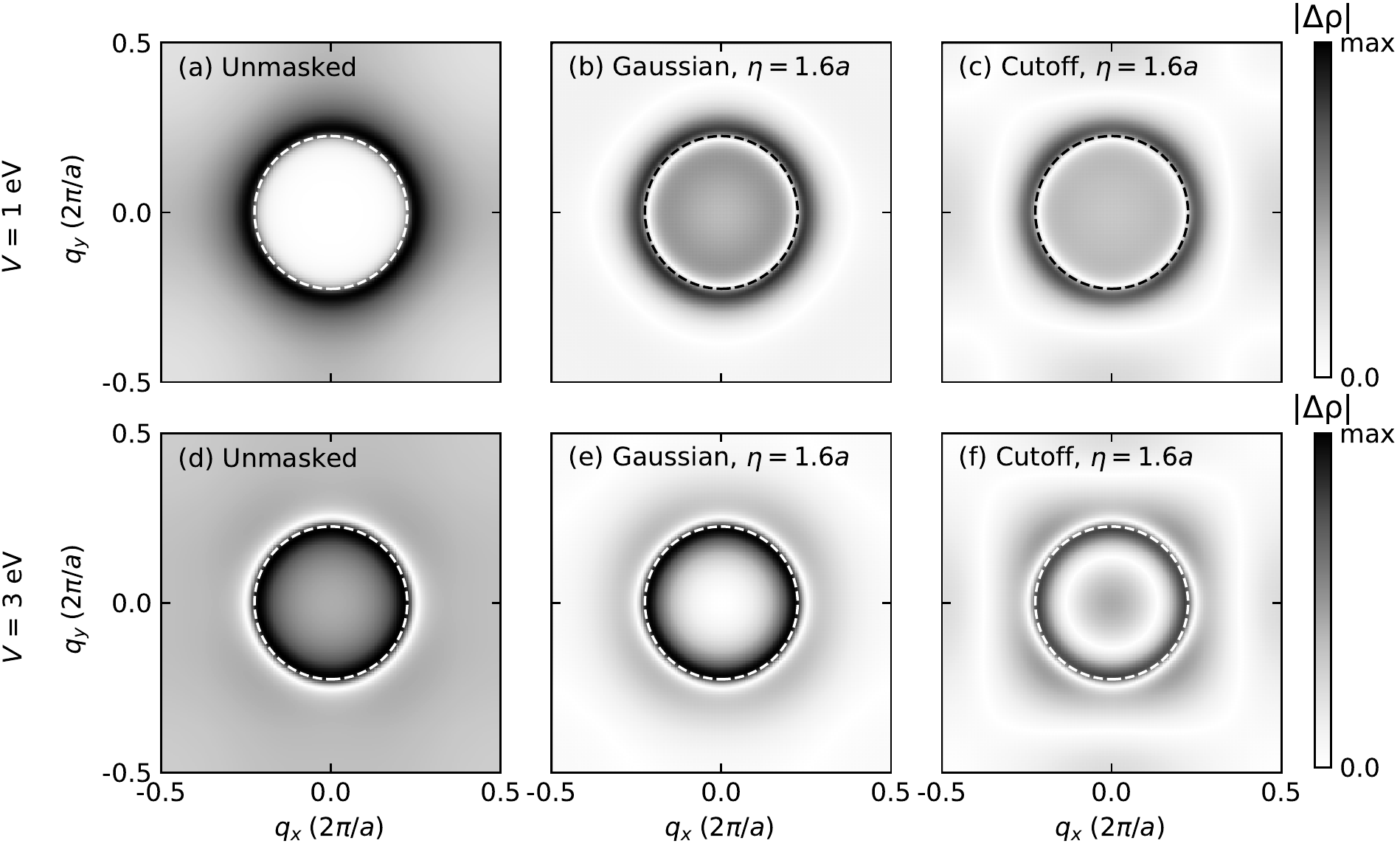}
    \caption{Effects of the real-space masking on the QPI patterns of the square lattice with $t=-1$~eV at $\omega=3.5$~eV. (a) and (d): Unmasked QPI patterns. (b) and (e): Gaussian masked QPI patterns, (c) and (f): Cutoff masked QPI patterns. The dashed curves are circles with radius $q=2k_\mathrm{c}$.}
    \label{fig:supp_mask}
\end{figure*}

Figure~\ref{fig:supp_mask} shows the effects of real-space masking on the QPI patterns. Figure~\ref{fig:supp_mask} (b), (c), (e), and (f) are obtained by Fourier transforming the masked LDOS variation $\Delta\rho_{\rm masked}({\bf R};\omega)$ instead of the unmasked one $\Delta\rho({\bf R};\omega)$.
For the Gaussian masking, the masked LDOS variation is
\begin{equation}
    \Delta\rho_{\rm masked}({\bf R};\omega)
    = \Delta\rho({\bf R};\omega) \times [1-\exp(-R^2/\eta^2)].
\end{equation}
For the cutoff masking, we set
\begin{equation}
    \Delta\rho_{\rm masked}({\bf R};\omega)
    = \Delta\rho({\bf R};\omega) \times \Theta(R-\eta)
\end{equation}
with $\Theta(x)$ the Heaviside step function.

\section{Details of First-Principles Calculations}
We used the \texttt{Quantum ESPRESSO} package~\cite{giannozzi_quantum_2009,giannozzi_advanced_2017} for the density functional theory computation with a plane-wave basis set.
The kinetic-energy cutoff of 70 Ry was used.
The parametrization of Perdew, Burke, and Ernzerhof (PBE) was used for the generalized-gradient approximation of the exchange-correlation functional~\cite{PhysRevLett.77.3865}.
We included spin-orbit coupling by using fully relativistic pseudopotentials.
Fully relativistic pseudopotentials for Ta and As were taken from the SG15 library~\cite{scherpelz_implementation_2016,PhysRevB.88.085117,Schlipf2015}.
Magnetism was not considered.
We applied the Methfessel-Paxton smearing~\cite{PhysRevB.40.3616} of 0.01~Ry. The k-point grid for the bulk calculation was set to $12\times12\times8$. For the slab calculation, we used a slab with 40 atomic layers and used a $12\times12\times1$ k-point grid.
We used the experimental lattice parameters of TaAs~\cite{furuseth1965arsenides}.
We relaxed the atomic coordinates of the bulk until the forces on the atoms were below $2\times 10^{-3}$~eV/$\mathrm{\AA}$.
We did not relax the structure of the slab.

To generate the Wannier-function-based tight-binding Hamiltonian, we used the Wannier90 package~\cite{mostofi_updated_2014,Pizzi2020}.
We used the projection-only Wannier functions, which are generated without any iterative localization procedures. Consequently, the bulk and slab tight-binding models could be stitched to a semi-infinite surface model without any corrections~\cite{PhysRevB.99.125117}.
For Wannierization, the Brillouin zone was sampled with uniform $7\times7\times6$ and $7\times7\times1$ grids for the bulk and slab, respectively.
We used the Ta-centered $d$ orbitals and As-centered $p$ orbitals as the initial guesses for the Wannier functions.
The spinor parts of the initial guesses were aligned along the $z$ axis.
The inner (frozen) windows were set to [-2, 1] eV around the Fermi level for the bulk and slab. The outer (disentanglement) windows were set to [-10, 9] eV and [-9.5, 9.5] eV around the Fermi level for the bulk and slab, respectively.
The ab initio tight-binding Hamiltonian is symmetrized by zeroing out a hopping integral if any of its symmetry-equivalent hopping integrals is not included in the tight-binding Hamiltonian due to the finite Brillouin zone sampling.

\begin{figure}[!tb]
    \centering
    \includegraphics[width=0.45\textwidth]{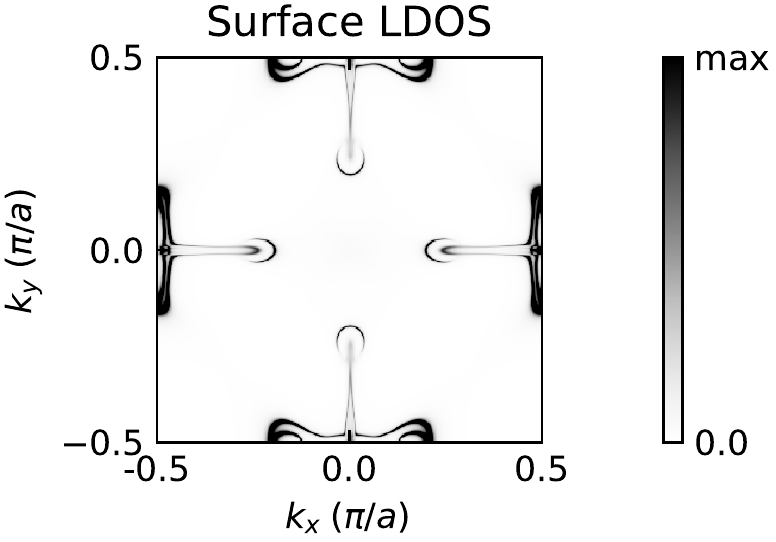}
    \caption{Fourier transform of the LDOS at the orbitals localized at the topmost As atom for the As-terminated TaAs surface at the Fermi level.}
    \label{fig:supp_TaAs_LDOS}
\end{figure}

In Fig.~\ref{fig:supp_TaAs_LDOS}, we plot the surface LDOS of the TaAs surface. There, the surface Fermi arcs are observed. The computed surface LDOS agrees with previous surface-specific calculations~\cite{2015SunTaAs,2015HuangTaAs,inoue_quasiparticle_2016}.

\begin{figure}[!tb]
    \centering
    \includegraphics[width=0.45  \textwidth]{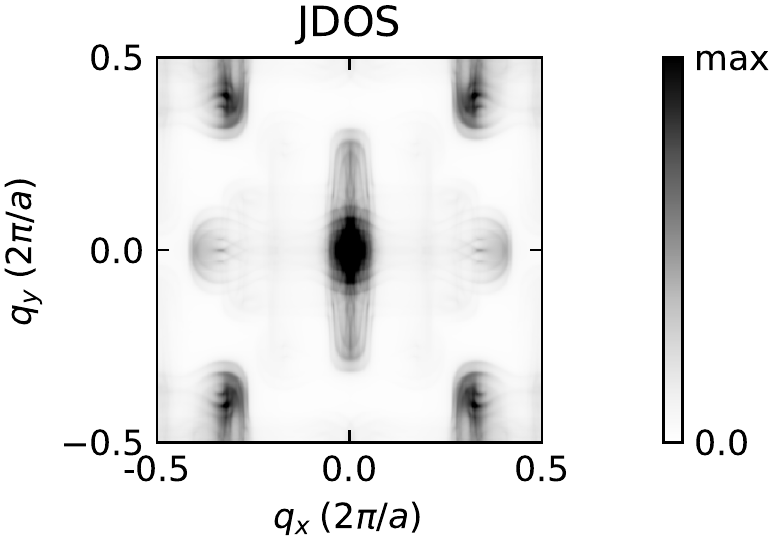}
    \caption{Surface-projected JDOS for the orbitals localized at the topmost As atom for the As-terminated TaAs surface for $\omega$ at the Fermi level.}
    \label{fig:supp_JDOS}
\end{figure}

\begin{figure*}[!tb]
    \centering
    \includegraphics[width=0.95\textwidth]{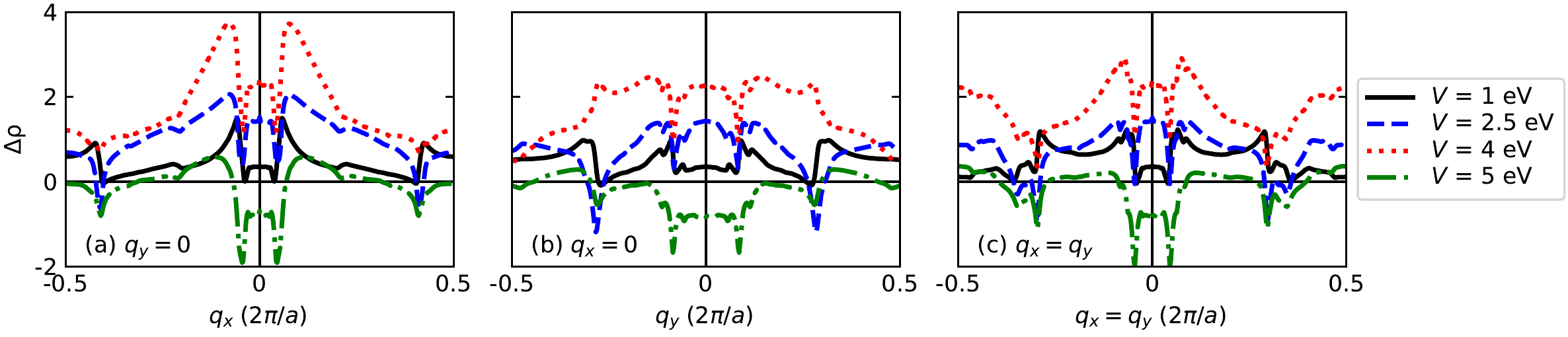}
    \caption{Signed values of the QPI signal for the As-terminated TaAs surface along the $q_y = 0$ [(a)], $q_x = 0$ [(b)], and $q_x = q_y$ [(c)] lines.}
    \label{fig:supp_cuts}
\end{figure*}

\begin{figure*}[!tb]
    \centering
    \includegraphics[width=\textwidth]{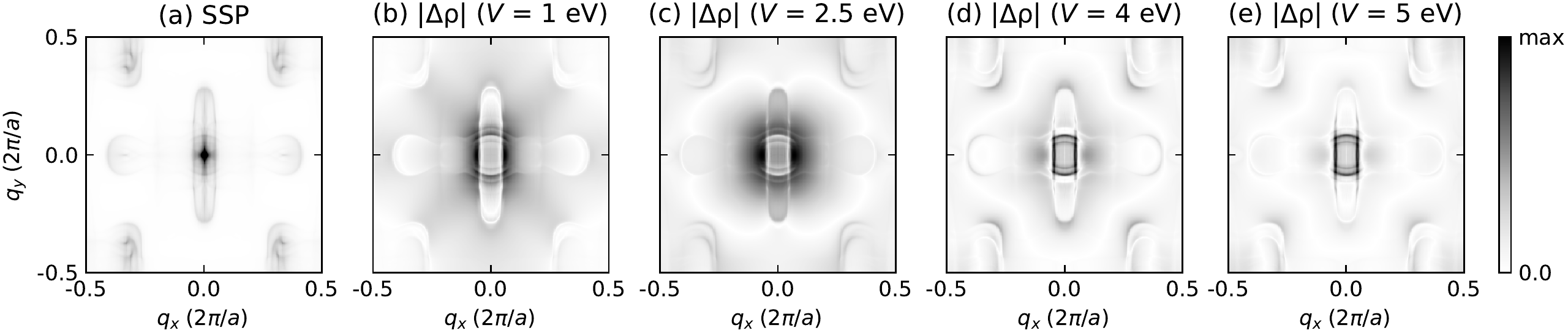}
    \caption{SSP [(a)], and the absolute value of the QPI patterns [(b)-(e)] for the As-terminated TaAs surface for $\omega$ at the Fermi level. Here, the on-site impurity potentials of strength $V$ and 0.5$V$ were applied at the $p$ orbitals of a single surface As atom and those of the nearest-neighboring As atoms, respectively.}
    \label{fig:nn_onsite}
\end{figure*}

\begin{acknowledgments}
\section{Acknowledgement}
$^\S$S.-J.H. and J.-M.L. contributed equally to this work.
\end{acknowledgments}

\newpage

\bibliography{supplement}